# Rapid Doubling of the Critical Current of $YBa_2Cu_3O_{7-\delta}$ Coated Conductors for Viable High-Speed Industrial Processing

M. Leroux[1], K. J. Kihlstrom[1,4], S. Holleis[1,5], M. W. Rupich[3], S. Sathyamurthy[3], S. Fleshler[3], H. P. Sheng[2], D. J. Miller[2], S. Eley[6], L. Civale[6], A. Kayani[7], P. M. Niraula[7], U. Welp[1] and W.-K. Kwok[1]

[1] *Materials Science Division, Argonne National Laboratory, Argonne, IL 60439, USA*
[2] *Electron Microscopy Center-Center for Nanoscale Materials, Argonne National Laboratory, Argonne, IL 60439, USA*
[3] *American Superconductor Corp., 64 Jackson Road, Devens, MA 01434, USA*
[4] *Department of Physics, University of Illinois at Chicago, Chicago, IL 60607, USA*
[5] *Atominstitut, TU Wien, Stadionallee 2, 1020 Vienna, Austria*
[6] *Materials Physics and Applications Division (MPA) & Condensed Matter and Magnet Science (CMMS), Los Alamos, NM 87545, USA*
[7] *Department of Physics, Western Michigan University, Kalamazoo, MI 49008, USA.*

We demonstrate that 3.5-MeV oxygen irradiation can markedly enhance the in-field critical current of commercial 2nd generation superconducting tapes with an exposure time of just one second per 0.8 cm$^2$. The speed demonstrated here is now at the level required for an industrial reel-to-reel post-processing. The irradiation is made on production line samples through the protective silver coating and does not require any modification of the growth process.   From TEM imaging, we identify small clusters as the main source of increased vortex pinning.

Increasing the current carrying capacity of 2$^{nd}$ generation (2G) YBCO high temperature superconducting (HTS) wires in the presence of high magnetic fields is critical for the commercialization of HTS based rotating machine applications such as lightweight and compact off-shore wind turbines and motors as well as various HTS magnet applications [1-4]. For these, operation in magnetic fields of several Tesla and at temperatures around 30K is envisioned. Although conductors of hundreds of meters in length with self-field critical current densities $J_c$ of more than 3 – 4 MA/cm$^2$ (more than 300 – 400 A/cm-width) at 77 K can now reliably be manufactured, the rapid suppression of $J_c$ in even modest applied magnetic fields continues to be a major challenge for HTS conductor development.

In recent years, impressive advances in the in-field performance of short-length samples have been achieved [5-10], largely due to the strict control over the micro- and nanostructures. The formation of the desired pinning centers depends sensitively on the film deposition technique and substrate architecture. For instance, self-assembled nanorods can be engineered in films grown by pulsed laser deposition (PLD) or MOCVD from material containing excess metal oxides such as $BaZrO_3$ [11-14], $BaSnO_3$ [15] or $BaHfO_3$ [16], whereas the deposition of films with excess Zr using metal organic deposition (MOD) on single-crystal substrates [17] and on IBAD substrates [18] does not yield nanorods but nanoparticles. In general, the enhanced vortex pinning arises from the complex combined effects of the introduced second phases (nanorods or nanoparticles), additional structural disorder such as twin boundaries, stacking faults and point defects, as well as from isotropic pinning due to strain fields [5, 17]. In short-length samples, critical current densities as high as ~ 7 MA/cm$^2$ at 30 K and 9 T applied parallel to c-axis have been reported [9]. The translation of these advances into a reliable large-scale production process is a time consuming process currently under development.

An alternative to increase the critical current density by modifying the chemical synthesis is afforded by particle irradiation, which may be applicable to all superconducting materials. Depending on the mass and energy of the ions and the properties of the superconducting material, irradiation enables the creation of defects with well-controlled density and topology, such as points, clusters or tracks. Their effect on vortex pinning in superconductors as diverse as HTS single crystals and epitaxial films [19-25] and Fe-based superconductors [26, 27] has been extensively studied. However, only recently has their potential for improving the performance of commercial coated conductors been explored [28-35]. In particular, in our previous work [29] we have demonstrated that the in-field critical current density of short samples cut from production-line coated conductors can be doubled using irradiation with 4 MeV protons. It is believed that the mixed pinning landscapes [14, 18, 29, 36-38] that arise from the superposition of pre-existing defect structures and ~5 nm sized irradiation-induced defects are particularly efficient for vortex pinning in high fields. Significant effort has been devoted to reducing the size of artificial pinning centers to the 5-nm level by chemical synthesis, but this requires modifications of the growth process and remains a challenge. In contrast, irradiation can easily produce defects of this size and is independent of the growth process. Furthermore, particle irradiation can create defect structures that are highly uniform along the conductor length [35], , an aspect that has proven a challenge for chemical synthesis routes.  Nevertheless, irradiation has so far been regarded as mostly of academic interest since the necessary irradiation times were orders of magnitude too long to be useful for industrial-scale production, and also because it required complex and expensive dedicated facilities.

Here we show that the irradiation of current production-line coated conductors with 3.5 MeV oxygen ions more than doubles their critical current density in high magnetic fields in a process that covers ~0.8 cm$^2$ of coated conductor per second. These results are achievable with a standard accelerator capable of delivering an ion beam current of 1 μA of $^{16}O^{3+}$ at a terminal voltage of 1.15 MV.  These specifications are fairly easily achieved, and ion-irradiation of coated conductors could join a global trend of new usage of particle accelerators in advanced manufacturing [39, 40].  Transmission electron microscopy (TEM) reveals that the irradiation-induced vortex pinning originates from a large number of nm-sized anisotropic defects uniformly dispersed throughout the sample.  From annealing studies (not shown) we find that the $J_c$-enhancement is stable up to ~200 $^o$C for over an hour, and thus should probably survive a quench.

Samples of Dy-doped YBCO conductors, 50 mm long and 11 mm wide, were cut from standard 46 mm wide production strips manufactured by American Superconductor Corp (AMSC). The superconducting films were deposited onto a Rolling Assisted Bi-axially Textured Substrate (RABiTS) using metal organic deposition MOD [41]. The current MOD growth process produces a uniform defect structure in the YBCO superconductor layer consisting of a dispersion of rare earth oxide nano-particles, stacking faults, twin boundaries and dislocations [42].

The samples were irradiated using the 6.0-MV tandem Van de Graaff accelerator at Western Michigan University with $^{16}O^{3+}$ ions beams oriented along the normal of the film plane (c-axis). Gold/carbon foils and collimators were used to shape the beam into a uniform round spot of 0.8 cm$^2$. In our study, beam energy of 4.6 MeV is required in order to compensate for the energy loss in the gold/carbon foils and achieve energy of 3.5 MeV on the sample. In a dedicated industrial accelerator, beam rasterizing could be employed instead of a foil to achieve uniformity, thus further reducing the terminal voltage requirements. The sample holder was cooled to -10 °C to prevent excessive heating of the samples by the beam [35].  Here we chose oxygen ions as irradiating species because they produce significantly more damage than protons, are innocuous to the YBCO material and high beam currents at relatively low terminal voltage (1.15 MV) are easily obtained.

The critical current density $J_c$ was determined from the magnetization hysteresis measured with a commercial SQUID magnetometer on 1.5x1.5 mm$^2$ square samples. The flux creep rates were obtained from the time decay of the hysteretic magnetization over a period of 1 hour. The squares were patterned using photolithography and Ar-ion milling at different positions along the length and width of the tape samples, and all showed robust and consistent superconducting properties.

The characterization of the defect structures was carried out using high-resolution and diffraction contrast TEM. The TEM samples were prepared by focused ion beam (FIB) lift-out methods followed by back-side low-energy Ar ion milling. This approach minimizes artifacts due to specimen preparation. Several TEM samples were prepared from different regions of each sample (irradiated and unirradiated control) to ensure that results were representative of each condition.

Simulations using the SRIM-TRIM software package [43] indicate that irradiation with 3.5 MeV oxygen ions creates approximately 10$^4$ times more displacements per incident particle than comparable 4 MeV proton irradiation [29] implying a significant reduction in irradiation time. Fig. 1 shows the damage profile created in the tape by oxygen ions for different energies. Most of the damage occurs within a few microns below the surface, in a limited range of depth at the end of the ion's flight path (Bragg peak). We conclude that a 3.5 MeV oxygen beam positions the Bragg peak inside the YBCO layer, thus increasing the defect creation in the middle of the YBCO layer 5-fold as compared to a 6.0 MeV oxygen beam; however, with a slightly less uniform through thickness damage profile. These estimates account for the energy loss of the ions in the silver cover layer. In contrast, previous studies using low-energy Au-ions considered bare YBCO films deposited on single crystal substrates [28].

The field-dependence of the critical current density, as obtained from the Bean critical state analysis of the magnetization hysteresis, is displayed in Fig. 2 on log-log scales before and after irradiation with 4.6-MeV oxygen ions (3.5 MeV on the sample) to a dose of 0.3x10$^{13}$ O/cm$^2$ together with the resulting enhancement factor $J_c$(after)/$J_c$(before). Absolute values of 3.5 MA/cm$^2$ at 27K-6T and 8.5 MA/cm$^2$ at 5K-6T are obtained for the irradiated samples. Fig. 2 also shows that the field-dependence of $J_c$ in fields above ~0.5 T is well approximated by a power law of the form $J_c \sim B^{-\alpha}$ and that the major effect of the irradiation is the reduction of the field-dependence of $J_c$, that is, the reduction of $\alpha$. Accordingly, the enhancement of $J_c$ is higher at high fields, nearly doubling near 6 T, whereas the effect in low fields is less. A similar evolution was found for proton irradiated [29], 75-MeV Ag irradiated [45] and for 18-MeV Au irradiated [35] coated conductors. The rapid reduction of $\alpha$ with irradiation dose from 0.75 in the pristine sample to about 0.5 at optimal dose and down to 0.4 for over-optimal dose is illustrated in Fig. 3a. These data will be discussed further below.

The dose dependence of the enhancement of $J_c$ after oxygen irradiation is presented in Fig. 3b. The optimal dose is field and temperature dependent: with decreasing temperature and increasing magnetic field, the peak enhancement requires a higher irradiation dose, similar to results on proton irradiated samples [44] and in agreement with recent simulations of vortex pinning in random arrays of spherical nanoparticles [46]. The dose-dependence data show a striking asymmetry in which the enhancement of $J_c$ rises rapidly at small doses, passes through the optimum and then decreases gradually at high doses. Such variation is beneficial for industrial application of the irradiation process since more than 80% of the enhancement is already obtained for $0.3 \times 10^{13}$ O-ion/cm$^2$ whereas optimum enhancement occurs at $0.6 \times 10^{13}$ O-ion/cm$^2$, that is, double the required time. Also of interest is the significant enhancement even at T = 45 K. We note though that there is no enhancement of $J_c$ due to oxygen ion irradiation at 77 K in fields applied along the c-axis. This finding is consistent with our earlier results on proton-irradiated coated conductors [29]. Furthermore, previous studies of irradiation effects in coated conductors due to 5-MeV Ni, 25-MeV-Au [30-32] and 75-MeV Ag ions [45] showed that $J_c$-enhancements at 77 K are generally small and strongly dependent on conductor composition and field-angle.

Fig. 4 shows diffraction contrast TEM images obtained with scattering vector (002) of a pristine (a) and a sample irradiated to a dose of $1 \times 10^{13}$ ions/cm$^2$ (b), respectively. The c-axis is horizontal and the layered structure of YBCO is clearly visible. The TEM images of the pristine sample reveal the defect structure expected for MOD coated conductors [30, 42]: mostly rare earth oxide nanoparticles, several tens of nanometer in diameter, and a few stacking faults. Twin boundaries and dislocations were not visible in the field of view under the imaging conditions in Fig. 4. Similar to proton irradiation, oxygen irradiation creates a large number of finely dispersed small defects approximately 5 nm in diameter.

The images in Fig. 4 suggest a model of a mixed pinning landscape composed of a large number of relatively small irradiation-induced defects coexisting with large pre-existing rare-earth oxide precipitates, twin boundaries, and point defects. At low fields, all vortices are assumed strongly pinned, $J_c$ is approximately field-independent and the irradiation-induced defects do not contribute to additional vortex pinning. With increasing field, $J_c$ is expected to initially vary approximately as $B^{-1/2}$ and, when interstitial vortices appear, as $B^{-1}$ [47-49]. Our data roughly follow this trend although $\alpha$ of the pristine sample does not reach the value of 1 at high fields, but instead acquires values of ~0.7 which are typical for a broad range of coated conductors [18]. With increasing field, the finely dispersed irradiation induced defects may effectively pin interstitial vortices accounting for the observed enhancement of $J_c$ in high fields. Furthermore, at this qualitative level, one could expect that the pinning of interstitial vortices becomes more efficient when more small defects are introduced, thus $\alpha$ should decrease with increasing irradiation dose, as seen in Fig. 3a.

The inset of Fig. 3a shows the temperature dependence of the normalized logarithmic relaxation rate S = dlnM/dlnt of the magnetization for oxygen and proton irradiated samples in a field of 1 T || c.  Here, M is the hysteretic magnetization of the sample, which is proportional to the supercurrent density.  Temperature-dependent creep rates, S(T), for a sample irradiated with protons at nearly the optimal dose ($8\times10^{16}$ p/cm$^2$), two samples irradiated with oxygen each at different doses ($0.3\times10^{13}$ O/cm$^2$ and $3\times10^{13}$ O/cm$^2$), and two un-irradiated reference samples are compared.  It is remarkable that the creep rate for all samples approaches a common, almost linear, temperature dependence at temperatures below 20 K.  A similar finding has been reported for coated conductors grown using different synthesis routes [50]. According to the classic Anderson-Kim model of vortex creep [51] a linear S(T)-relation of the form S = $k_B T/U_0$ is expected, where $U_0$ is the dominant vortex pinning potential.  From the data in Fig. 3a, one estimates $U_0$ ~ 1200 K for all samples, independent of the irradiation state suggesting that vortex dynamics is dominated by strong pre-existing pinning, i.e. rare earth nanoparticles and twin boundaries.  One also notes that S(T) does not extrapolate to zero, indicative of a contribution due to quantum creep at very low temperatures.  Above 20 K, all samples show a clear departure from the almost linear S(T) dependence, pointing to a fundamental change in the vortex dynamics.  S(T) of the pristine sample is non-monotonic with a minimum around 40 K and the creep rate strongly increases with irradiation dose.  This evolution is suggestive of a crossover from mostly strong single-vortex pinning at low fields and low temperatures to mostly collective pinning at high fields and high temperatures. Coincidentally, the temperature dependence of $J_c$ of coated conductors in a field of several Tesla displays a cross-over around 25 K from a regime of steep decrease of $J_c$(T) at low temperatures to a regime of weaker decrease at high temperatures [52].  A minimum in S(T) has also been observed in other YBCO films with pinning dominated by large nanoparticles and is likely due to the large pinning energy of these defects [50].  The creep rate in this temperature range strongly increases with both oxygen and proton irradiation dose, consistent with a scenario where pinning becomes progressively dominated by smaller defects.

In summary, we have demonstrated the doubling of the critical current density of state-of-the-art production-line REBCO coated conductors in fields of 6 T || c at 27 K, using a 3.5 MeV oxygen beam with an exposure time of the order of one second. This doubling of $J_c$ within one second or less, opens an industrially viable approach to address the challenge in HTS conductor development, namely their greatly reduced performance in even modest applied magnetic fields.  TEM images reveal that the enhanced critical current is due to finely dispersed small clusters approximately 5 nm in diameter and with anisotropic shape.  The major effect of the irradiation-induced defects is the reduction of the field-dependence of $J_c$, which we attribute to the mixed pinning landscape composed of strong pre-existing pin sites and the finely dispersed irradiation-induced defects.

Acknowledgements: This work was supported as part of the Center for Emergent Superconductivity, an Energy Frontier Research Center funded by the U.S. Department of Energy, Office of Science, Office of Basic Energy Sciences. Patterning of the samples and their microstructural characterization were performed at the Center for Nanoscale Materials. Use of the Center for Nanoscale Materials, an Office of Science user facility, was supported by the U. S. Department of Energy, Office of Science, Office of Basic Energy Sciences, under Contract No. DE-AC02-06CH11357. SH acknowledges a TOP Stipendium Niederoesterreich and support from the KUWI program at the TU Wien. ML acknowledges J. Greene of the Physics Division Accelerator Target Laboratory at Argonne National Laboratory for providing the composite Au/C targets used in this study.

References

1) A. P. Malozemoff, Annu. Rev. Mater. Res. **42,** 373–397 (2012).
2) Y. Shiohara, T. Taneda, and M. Yoshizumi, Jpn. J. Appl. Phys. **51,** 010007 (2012).
3) C. Senatore, *et al.*, Supercond. Sci. Technol. **27,** 103001 (2014).
4) X. Obradors, T. Puig, Supercond. Sci. Technol. **27,** 044003 (2014).
5) A. Xu, V. Braccini, J. Jaroszynski, Y. Xin, D. C. Larbalestier, Phys. Rev. B **86,** 115416 (2012).
6) V. Selvamanickam, V. *et al.*, Supercond. Sci. Technol. **26,** 035006 (2013).
7) A. Xu, L. Delgado, N, Khatri, Y. Lin, V. Selvamanickam, D. Abraimov, J. Jaroszynski, F. Kametami, D. C. Larblalestier, APL Materials **2**, 046111 (2014).
8) V. Selvamanickam, M. H. Gharahcheshmeh, A. Xu, Y. Zhang, E. Galstyan, Supercond. Sci. Technol. **28**, 072002 (2015).
9) V. Selvamanickam, *et al.,* Appl. Phys. Lett. **106,** 032601 (2015).
10) S. Awaji, Y. Yoshida, T. Suzuki, K. Watanabe, K. Hikawa, Y. Ichino, T. Izumi, Applied Physics Express **8**, 023101 (2015).
11) J. L. MacManus-Driscoll et al., Nature Mater. **3**, 439–443 (2004).
12) T. Haugan et al., Nature **430**, 867 (2004).
13) S. Kang et al., Science **311**, 1911 (2006).
14) B. Maiorov et al., Nature Mater. **8**, 398 (2009).
15) P. Mele, K. Matsumoto, T. Horide, A. Ichinose, M. Mukaida, Y. Yoshida, S. Horii, R. Kita, Supercond. Sci. Technol. **21**, 032002 (2008).
16) H. Tobita, K. Notoh, K. Higashikawa, M. Inoue, T. Kiss, T. Kato, T. Hirayama, M. Yoshizumi, T. Izumi, Y. Shiohara, Supercond. Sci. Technol. **25**, 062002 (2012).
17) J. Guiterrez, A. Llordes, J. Gazquez, M. Gibert, N. Roma, S. Ricart, A. Pomar, F. Sandiumenge, N. Mestres, T. Puig, X. Obradors, Nature Materials **6**, 367 (2007).
18) M. Miura, B. Maiorov, S. A. Baily, N. Haberkorn, J. O. Willis, K. Marken, T. Izumi, Y. Shiohara, L. Civale, Phys. Rev. B **83**, 184519 (2011).
19) B. Roas, B. Hensel , G. Saemann-Ischenko, L. Schultz, Appl. Phys. Lett. **54**, 1051 (1989).
20) L. Civale, A. D. Marwick, M. W. McElfresh, T. K. Worthington, A. P. Malozemofl, F. H. Holtzberg, J. R. Thomson, M. A. Kirk, Phys. Rev. Lett. **65**, 1164 (1990).
21) L. Civale et al., Phys. Rev. Lett. **67**, 648 (1991).
22) Y. Zhu, Z. X. Cai, R. C. Budhani, M. Suenaga, D. O. Welch, Phys. Rev. B **48**, 6436 (1993).
23) L. Civale, Supercond. Sci. Technol. **10**, A11–A28 (1997).
24) W. K. Kwok, L. M. Paulius, V. M. Vinokur, A. M. Petrean, R. M. Ronningen, G. W. Crabtree, Phys. Rev. B **58**, 14594 (1998).
25) J. Hua, U. Welp, J. Schlueter, A. Kayani, Z. L. Xiao, G. W. Cratree, W. K. Kwok Phys. Rev. B **82**, 024505 (2010).
26) T. Tamegai, T. Taen, H. Yagyuda,Y. Tsuchiya, S. Mohan, T. Taniguchi, Y. Nakajima, S. Okayasu, M. Sasase, H. Kitamura et al., Supercond. Sci. Technol. **25**, 084008 (2012).

27) L. Fang, Y. Jia, V. Mishra, C. Chaparro, V. K. Vlasko-Vlasov, A. E. Koshelev, U. Welp, G. W. Crabtree, S. Zhu, N. D. Zhigadlo, S. Katrych, J. Karpinski, W. K. Kwok, Nature Commun. **4**, 2655 (2013).
28) H. Matsui, H. Ogiso, H. Yamasaki, T. Kumagai, M. Sohma, I. Yamaguchi, T. Manabe, Appl. Phys. Lett. **101**, 232601 (2012).
29) Y. Jia, M. LeRoux, D. J. Miller, J. G. Wen, W. K. Kwok, U. Welp, M. W. Rupich, X. Li, S. Sathyamurthy, S. Fleshler, A. P. Malozemoff, A. Kayani, O. Ayala-Valenzuela, L. Civale , Appl. Phys. Lett. **103**, (2013).
30) K. J. Leonard, T. Aytug, F. A. List, III, A. Perez-Bergquist, W. J. Weber, A. Gapud, Fusion Reactor Materials Program, June 30, 2013, DOE/ER-0313/54, Vol. **54**, p. 125.
31) K. J. Leonard, T. Aytug, F. A. List, III, A. Perez-Bergquist, W. J. Weber, A. Gapud, Fusion Reactor Materials Program, December 31, 2013, DOE/ER-0313/55, Vol. **55**, p. 54.
32) K. J. Leonard, T. Aytug, A. A. Gapud, F. A. List III, N. T. Greenwood, Y. W. Zhang, A. G. Perez-Bergquist, W. J. Weber, Fusion Science and Technology **66**, 57 (2014).
33) H. Matsui, H. Ogiso, H. Yamasaki, M. Sohma, I. Yamaguchi, T. Kumagai, T. Manabe, Journal of Physics: Conference Series **507**, 022019 (2014).
34) H. Matsui, T. Ootsuka, H. Ogiso, H. Yamasaki, M. Sohma, I. Yamaguchi, T. Kumagai, T. Manabe, J. Appl. Phys. **117**, 043911 (2015).
35) M. W. Rupich, S. Sathyamurthy, S. Fleshler, Q. Li, V. Solovyov, T. Ozaki, U. Welp, W.-K. Kwok, M. Leroux, A. E. Koshelev, D. J. Miller, K. Kihlstrom, L. Civale, S. Eley, A. Kayani, Engineered Pinning Landscapes for Enhanced 2G Coil Wire, IEEE Trans. Appl. Superconductivity, EUCAS 2015.
36) T. G. Holesinger et al., Adv. Mater. **20**, 391–407 (2008).
37) H. Yamasaki, K. Ohki, H. Yamada, Y. Nakagawa, and Y. Mawatari, Supercond. Sci. Technol. **21**, 125011 (2008).
38) K. J. Kihlstrom, L. Fang, Y. Jia, B. Shen, A. E. Koshelev, U. Welp, G. W. Crabtree, W.-K. Kwok, A. Kayani, S. F. Zhu, H.-H. Wen, Appl. Phys. Lett. **103**, 202601 (2013).
39) T. Feder, Phys. Today **68**, 18 (2015).
40) R. W. Hamm, M. E. Hamm, Phys. Today **64**, 46 (2011).
41) M. Rupich, D. Verebelyi, W. Zhang, T. Kodenkandath, and X. Li, MRS Bull. **29**, 572 (2004); X. Li, M. W. Rupich, C. L. H. Thieme, M. Teplitsky, S. Sathyamurthy, E. Thompson, E. Siegal, D. Buczek, J. Schreiber, K. DeMoranville et al., IEEE Trans. Appl. Supercond. **19**, 3231 (2009).
42) J. A. Xia, N. Long, N. Strickland, P. Hoefakker, E. F. Talantsev, X. Li, W. Zhang, T. Kodenkandath, Y. Huang, M. Rupich, Supercond. Sci. Technol. **20**, 880 (2007).
43) J. F. Ziegler, J. P. Biersack, M. D. Ziegler, SRIM, The Stopping and Range of Ions in Matter.
44) M. Leroux, unpublished.
45) N. M. Strickland, E. F. Talantsev, N. J. Long, J. A. Xia, S. D. Searle, J. Kennedy, A. Markwitz, M. W. Rupich, X. Li, S. Sathyamurthy, Physica C **469**, 2060 (2009).
46) A. E. Koshelev, I. A. Sadovskyy, C. L. Phillips, and A. Glatz, arXiv:1509.04212
47) Y. Ovchinnikov, B. Ivlev, Phys. Rev. B **43**, 8024–8029 (1991).
48) A. E. Koshelev, A. B. Kolton, Phys. Rev. B **84**, 104528 (2011).
49) C. van der Beek et al., Phys. Rev. B **66**, 024523 (2002).

50) N. Haberkorn, M. Miura, J. Baca, B. Maiorov, I. Usov, P. Dowden, S. R. Foltyn, T. G. Holesinger, J. O. Willis, K. R. Marken, T. Izumi, Y. Shiohara, L. Civale, Phys. Rev. B **85**, 174504 (2012).
51) Y. Yeshurun, A. P. Malozemoff, A. Shaulov, Reviews of Modern Physics **68**, 911 (1996).
52) M. W. Rupich, 2G HTS Wire Performance and R&D Focus at AMSC, Presented at the 2014 Applied Superconductivity Conference, August 14, 2014, Charlotte, NC.

# Figures

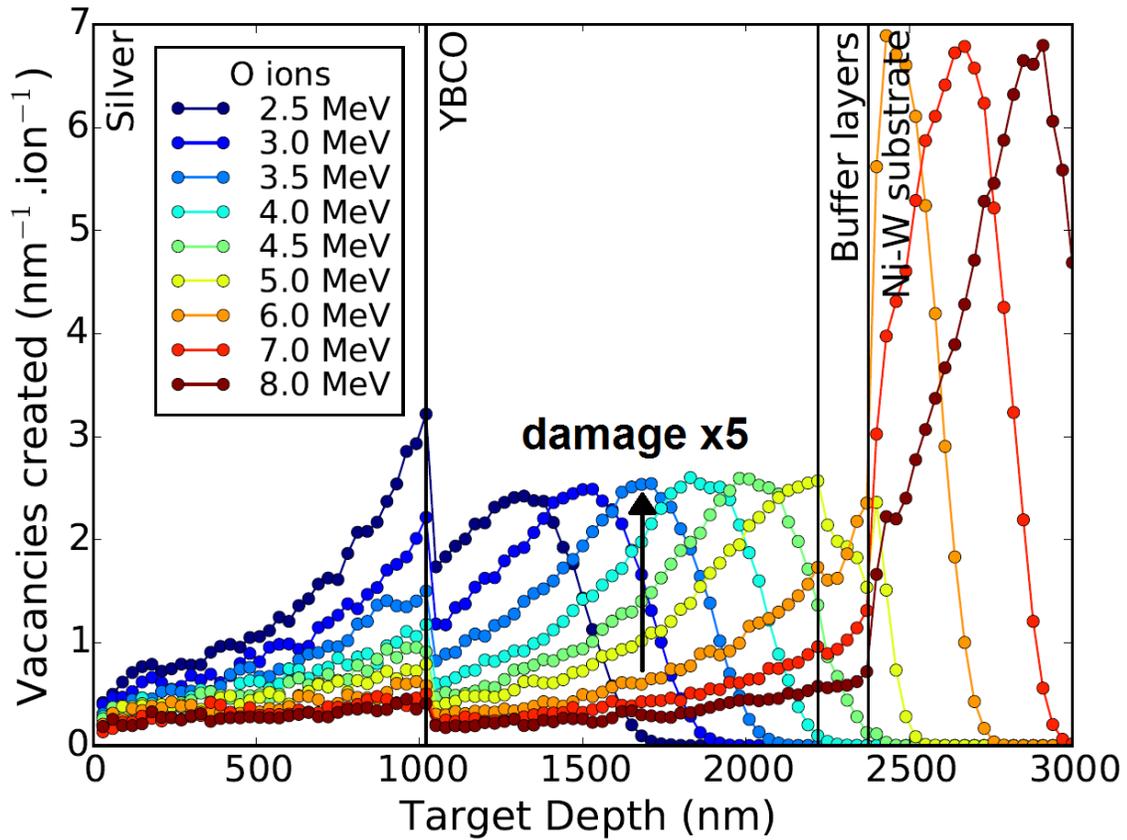

Fig. 1 SRIM-TRIM simulations showing the number of vacancies created by oxygen irradiation in the YBCO tape layers. For higher incident beam energies, the peak of maximum damage, or Bragg peak, occurs deeper in the material. While at 6.0 MeV the peak occurs in the substrate and only the tail contributes to creating defects in the YBCO layer, at 3.5 MeV the peak is in the YBCO layer thus enhancing the damage creation rate by a factor of 5 in the middle of the layer.

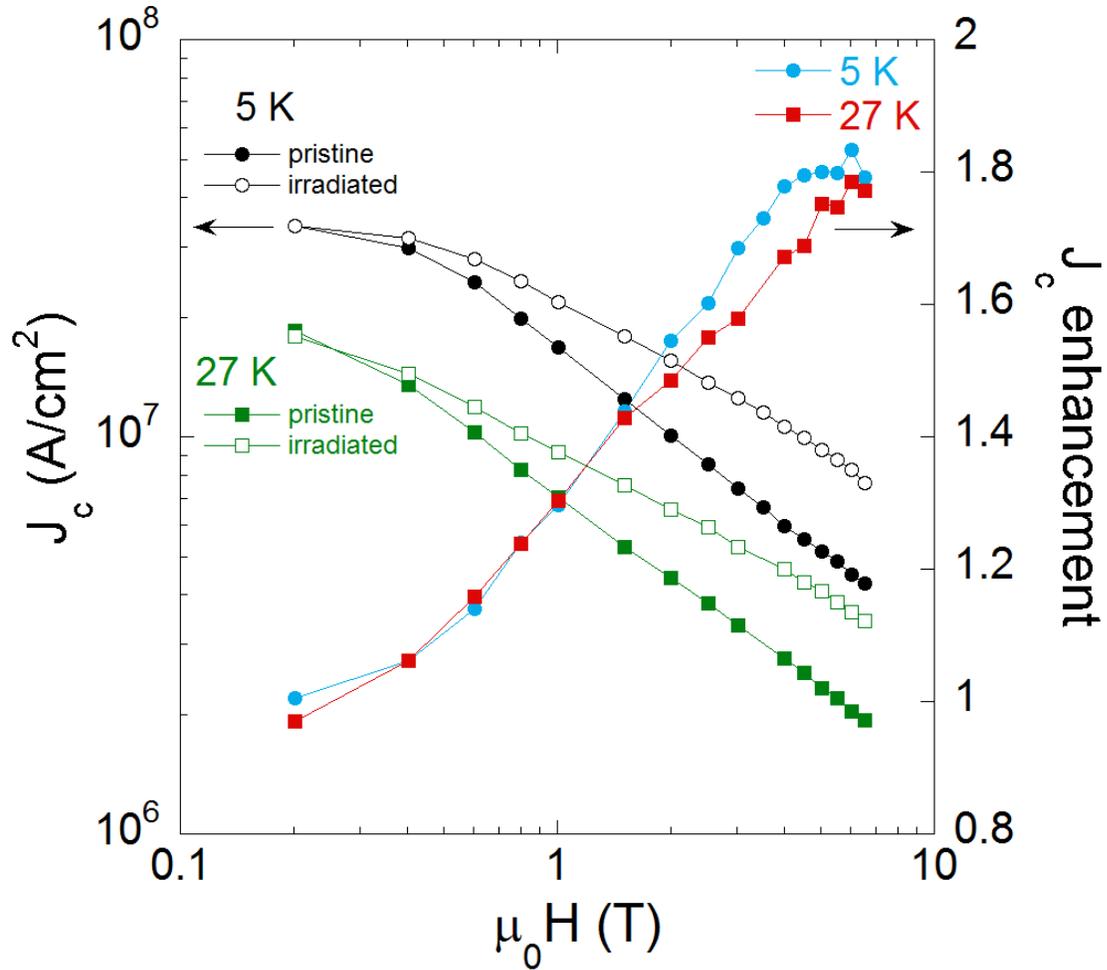

Fig. 2 Magnetic field dependence at 5 K and 27 K of the critical current density of superconducting coated conductor samples from AMSC both before and after oxygen irradiation at 3.5 MeV to a dose of 0.3x10$^{13}$ O-ion/cm$^2$. The data at high field is well described by $J_c = J_{c0} \times B^{-\alpha}$. The enhancement factor $J_c$(after)/$J_c$(before) is shown on the right y-axis. In high field, the critical current is nearly doubled.

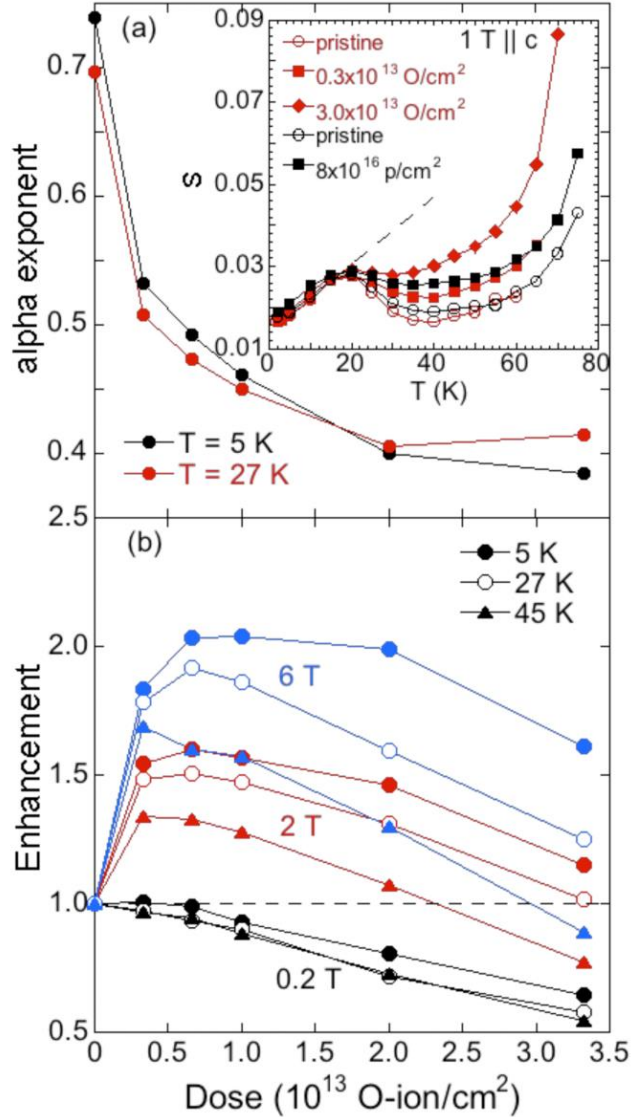

Fig 3 a) Dose dependence of the exponent $\alpha$ in $J_c = J_{c0} \times B^{-\alpha}$, for 3.5 MeV oxygen irradiation. The inset shows a comparison of the normalized logarithmic relaxation rate S of the magnetization as a function of temperature, before and after irradiation with 3.5 MeV oxygen ions (red symbols) and with 4 MeV protons (black symbols), in a field of 1 T ||c. The open red and black circles represent the data on the respective pristine samples, whereas the red and black squares correspond to near optimum irradiation doses. The dashed line indicates the common slope of the S(T)-curves at low temperatures. (b) Irradiation dose dependence of the enhancement of the critical current at temperatures of 5, 27 and 45 K (identified by open and closed circles and closed triangles) in fields of 0.2, 2 and 6 T (represented by black, red and blue colors). The optimal irradiation dose ranges from 0.3 to $1 \times 10^{13}$ O-ions/cm$^2$ depending on the temperature and field. The higher the field and the lower the temperature, the higher is the optimal irradiation dose.

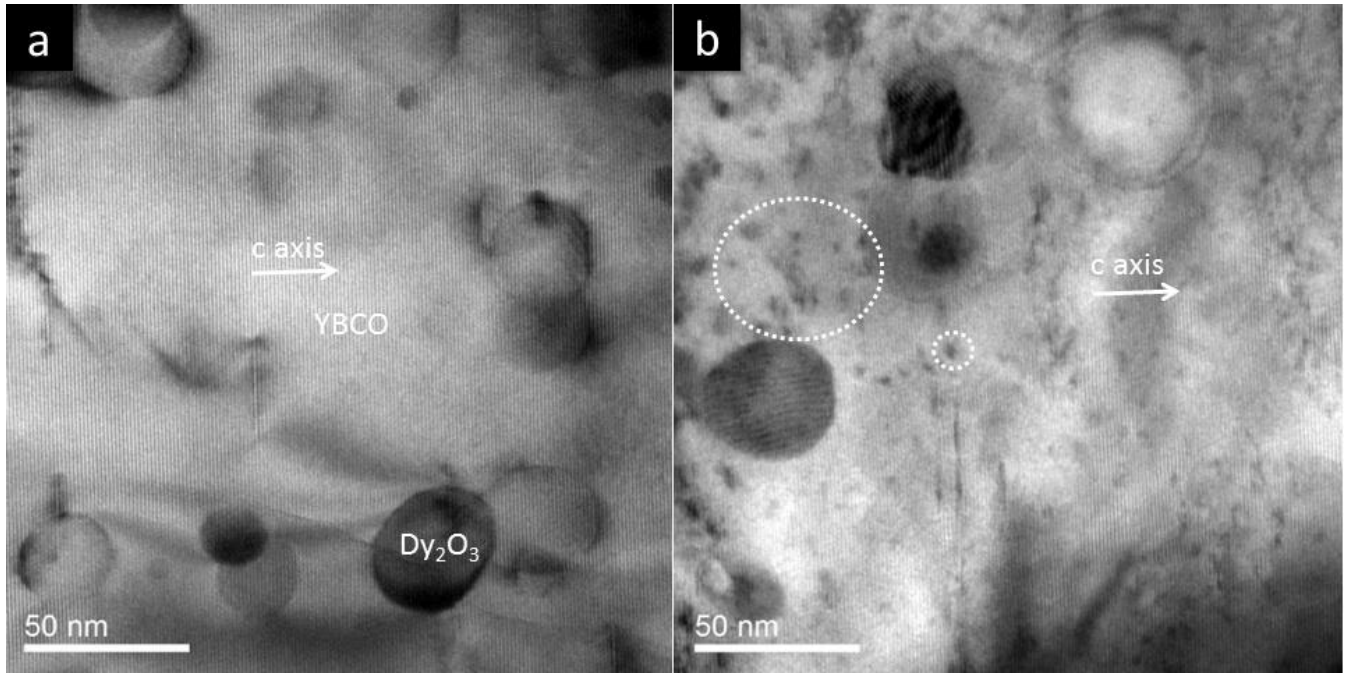

Fig. 4 Diffraction contrast TEM images of pristine (a) and irradiated (b) samples for diffraction vector (002). In the pristine samples the TEM baseline structure shows large $Dy_2O_3$ nanoparticles (~20-50 nm), stacking faults and large defect-free regions, whereas in the irradiated samples small defects ~5 nm in diameter appear randomly distributed within the baseline structure as highlighted by the dashed white circles.